\documentclass{aa}

\usepackage{hyperref}
\usepackage{graphicx}
\usepackage{natbib  }
\usepackage{color   }
\usepackage{array   }
\usepackage{txfonts }
\usepackage{breakurl}
\usepackage{multirow}
\usepackage{multicol}
\usepackage{booktabs}
\usepackage{dcolumn }
\usepackage{amstext }

\def\kms{km~s$^{-1}$}                    
\def\am{$^\prime$}                       
\def\as{$^{\prime\prime}$}               
\def\Msun{$\mathrm{M}_\odot$}            
\def\kmps{km~s$^{-1}$}                   
\newcommand{\cm}[1]{$\mathrm{cm}^{#1}$}  
\newcommand{\dms}[3]{$#1^{#2}\!\!.#3$}   
\def\twelveCO{$^{12}\mathrm{CO}$}        
\def\thirtyCO{$^{13}\mathrm{CO}$}        
\def\Xco{$X_\mathrm{CO}$}                
\def\COtrans{$J=1-0$}                    
\def\Fermi{{\itshape Fermi}-LAT}         
\def\Spitzer{{\itshape Spitzer}}         
\def\kes{Kes~41}                         

\begin{document} 

\title{Natal molecular cloud of SNR~{\kes}. Complete characterisation}

\titlerunning{Molecular and atomic line emission from the $\gamma$-ray emitting cloud towards SNR~{\kes}}

\author{L. Supan\inst{1,2}
   \and G. Castelletti \inst{1,2}
   \and A. D. Supanitsky \inst{1,2}
   \and M. G. Burton \inst{3,4}
   \and G. F. Wong \inst{3,5}
   \and C. Braiding \inst{3}}

\authorrunning{Supan et al.}

\offprints{L. Supan}
\institute {CONICET-Universidad de Buenos Aires, Instituto de Astronomía y Física del Espacio (IAFE), Buenos Aires, Argentina
\and Universidad de Buenos Aires, Facultad de Ciencias Exactas y Naturales. Buenos Aires, Argentina \\
     \email{lsupan@iafe.uba.ar}
\and School of Physics, University of New South Wales, Sydney, NSW 2052, Australia 
\and Armagh Observatory and Planetarium, College Hill, Armagh BT61 9DG, UK 
\and School of Computing Enginnering and Mathematics, Western Sydney University, Locked Bay 1797, Penrith, NSW 2751, Australia}

\date{Received 6 April 2018 / Accepted 13 August 2018}

\abstract
{Using high-resolution data of the $^{12}$CO and $^{13}$CO ({\COtrans}) line emission from The Mopra Southern Galactic Plane CO Survey in conjunction with neutral hydrogen observations from the Southern Galactic Plane Survey (SGPS) and mid-infrared {\Spitzer} data, we have explored the large-scale environment of the supernova remnant {\kes}. 
On the basis of these data, we identified for the first time the parent cloud of {\kes} in its whole extension and surveyed the HII regions, masers, and the population of massive young stellar objects in the cloud. 
The whole unveiled giant cloud, located at the kinematic distance of 12.0$\pm$3.6~kpc, whose average total mass and size are $\sim$10-30~$\times10^5$~{\Msun} and $\sim$$26^{\prime}$, also shines in $\gamma$-rays, as revealed by the Large Area Telescope on board the {\itshape Fermi} satellite. 
We determined a high average proton density $\sim$500-1000~\cm{-3} in the large molecular complex, of which protons from the neutral atomic and ionised gases comprise only $\sim$15\%.}


\maketitle

\section{Introduction}

Massive stars as well as their stellar remnants created after supernova (SN) events spatially coexist in their parent molecular cloud (MC) \citep{mck07}. 
The accelerated particles created at the SN shock fronts and/or in star-forming regions (SFRs) can cause emission at $\gamma$-ray energies as the result of their interaction with ambient photons and/or with the interstellar medium (ISM) in which these objects evolve \citep{banik+17, rom10}. 
Knowledge of the physical conditions of the ambient matter in bright $\gamma$-ray emitting molecular complexes is a particular important key in order to obtain information on the creation and distribution of Galactic cosmic rays. In these regards, surveys designed to cover the large-scale distribution of different molecular species and of atomic hydrogen \citep{hey15, kal09} constitute an ideal framework to assess these matters.

We here present a revisited analysis of the environs of the supernova remnant (SNR) {\kes}, where an excess in $\gamma$ rays at GeV energies was detected with the {\itshape Fermi}-Large Area Telescope (LAT) \citep{liu+15}. The interaction of {\kes} with nearby molecular gas has been known since the 1990s with the discovery of OH-1720~MHz maser emission \citep{koralesky+98}. A small portion of this gas about $\sim$~8$^{\prime}$ in size that also has a radial velocity of $\sim$$-$50~km~s$^{-1}$ was investigated by \citet{zhang+15}. Now, based on the interesting results reported by these authors, we have begun the analysis of the ISM in a significatively larger region using new information of the molecular $^{12}$CO and $^{13}$CO lines emission from The Mopra Southern Galactic Plane CO Survey, as well as neutral hydrogen and mid-infrared data from the Southern Galactic Plane Survey (SGPS) and {\Spitzer}, respectively. From our study we identified for the first time the giant parent molecular complex where {\kes}, a number of massive young stellar objects, and star-forming regions are mixed.

The paper is organised as follows: Sect.~\ref{data} describes the observations, and the complete characterisation of the discovered cloud along with the proton content determination are presented in Sect.~\ref{counterparts}. Sect.~\ref{summary} summarises our findings. The implications of the current study on the production of the observed $\gamma$ rays along with the modelling of the broadband emission from radio to $\gamma$-ray energies to understand the relative contribution of leptons and hadrons in the {\kes} region will be addressed in a companion paper (Supan et al. 2018b).

\section{Observations and data analysis}
\label{data}

\subsection{$^{12}$CO and $^{13}$CO($J$ = 1 $-$ 0) observations from The Mopra Southern Galactic Plane CO Survey}\label{data-co}
We used observations of the {\twelveCO} and {\thirtyCO} {\COtrans} rotational transition taken from The Mopra Southern Galactic Plane CO Survey \citep[hereafter, CO Mopra survey;][]{burton+13} to study the large-scale molecular structures in the region of SNR~{\kes}. For both CO isotopologues, the half-power beam width (HPBW) is 36{\as} at the observing frequencies, while the velocity resolution in each dataset is $\sim$0.09~{\kms}. Compared with the earlier CO survey of \citet{dam01}, the spatial and spectral resolution of the new Mopra data is about 14 times better in the region in which we are interested.

Both the {\twelveCO} and {\thirtyCO} ({\COtrans}) lines are important to obtain a compressive view of the molecular material. While the optically thick emission from the {\twelveCO} is in general significantly more intense than that observed in the {\thirtyCO} line, the latter is a more faithful tracer of the denser regions in the molecular gas. Therefore, we used both lines to determine the physical parameters of the molecular gas, in the way explained as follows. First, we computed the integrated emission of the {\twelveCO} in a region $R$ and in a given velocity range $\Delta v$ from the relation 

\begin{equation}\label{W12CO}
  W_{^{12}\mathrm{CO}} = \frac{1}{\eta_\mathrm{XB}} \iint_R \int_{\Delta v} T_{^{12}\mathrm{CO}}(l,b,v) \, dl \, db \, dv,
\end{equation}

\noindent
where $T_{^{12}\mathrm{CO}}$ corresponds to the brightness temperature of the {\twelveCO} emission. The factor $\eta_\mathrm{XB}$ is included as a correction made to account for the extended beam efficiency of the Mopra telescope, which at the observing frequency for the {\twelveCO} (115.27~GHz) is $\eta_\mathrm{XB}$ = 0.55 \citep{ladd+05}. 
Then, we calculated the integrated column density of the H$_{2}$ directly by adopting the CO-to-H$_{2}$ conversion factor {\Xco} = $N(\mathrm{H}_2) / W_{^{12}\mathrm{CO}}$ = $2.0\times10^{20}$~\cm{-2}~(K~\kmps)$^{-1}$, which has an estimated uncertainty of about 30\% \citep{bolatto+13}. 
On the other hand, in the case of  {\thirtyCO}, under the hypothesis of local thermodynamic equilibrium, it is possible to calculate the total column density $N(^{13}\mathrm{CO})$ according to

\noindent
\begin{equation}\label{N13CO}
  N(^{13}\mathrm{CO}) = \frac{1}{\eta_\mathrm{XB}} 2.42 \times 10^{14} \frac{T_\mathrm{ex}+0.88}{1-e^{5.29/{T_\mathrm{ex}}}} \iint_R \int_{\Delta v} \tau_{13}\, dl \, db \, dv ,
\end{equation}

\noindent
where $T_\mathrm{ex}$ represents the excitation temperature of the {\COtrans} transition and $\tau_{13} = \tau_{13}(l,b,v)$ is the corresponding optical depth of the {\thirtyCO} ({\COtrans}) spectral line \citep{wilson+13}. For this line (observed at 110.20~GHz), the corresponding correction is $\eta_\mathrm{XB}$ = 0.55 \citep{ladd+05}. 
Under the assumption that the {\thirtyCO} line is optically thin, the following approximation holds \citep{wilson+13}:

\noindent
\begin{equation}\label{tau13}
  \int_{\Delta v} \tau_{13} dv \approx \frac{1}{J(T_\mathrm{ex})-J(T_\mathrm{bkg})} \int_{\Delta v} T_{^{13}\mathrm{CO}}(v) \, dv ,
\end{equation}

\noindent
and the radiation temperature $J(T)$ is defined as

\begin{equation}\label{J}
  J(T) = \frac{5.29}{e^{5.29/T}-1} .
\end{equation}

In Eq.~\ref{tau13}, $T_{^{13}\mathrm{CO}}(v)$ denotes the brightness temperature of the emission for a gas moving at the radial velocity $v$, while $T_\mathrm{bkg}$ indicates the temperature of the background, $\sim$2.7~K. After we calculated the {\thirtyCO} column density, $N(\mathrm{H}_2)$ was determined using the conversion factor $N(^{13}\mathrm{CO}) / N(\mathrm{H}_2) = 5.6 \times 10^5$ given by \citet{simon+01}.

\subsection{HI observations}
\label{data-hi}
The emission of the neutral hydrogen (HI) 21 cm line was extracted from the Southern Galactic Plane Survey \citep[SGPS,][]{mcclure+05}, which combines observations carried out with the interferometer Australia Telescope Compact Array (ATCA) and the 64 m single-dish telescope of Parkes. These data have an angular resolution of \dms{2}{\prime}{2} and a separation between consecutive velocity channels of $\sim$0.8~{\kms}. The rms noise per velocity channel is $\sim$1.6~K.

The atomic contribution to the proton content of the medium can be calculated in the optically thin limit by integrating the HI emission in a velocity range $\Delta v$, using the expression \citep{dickey-lockman-90}

\begin{eqnarray}\label{NHI}
  N(\mathrm{HI}) &=& 1.823 \times 10^{18} \iint_R \int_{\Delta v} T(l,b,v)\, dl \, db \, dv ,
\end{eqnarray}

\noindent
where $T(l,b,v)$ is the brightness temperature of the emission of the HI gas moving at a radial velocity $v$ along the $(l,b)$ direction, within a region $R$.

\section{Results}
\label{counterparts}

\begin{figure*}[!ht]
 \centering
 \includegraphics[width=0.75\textwidth]{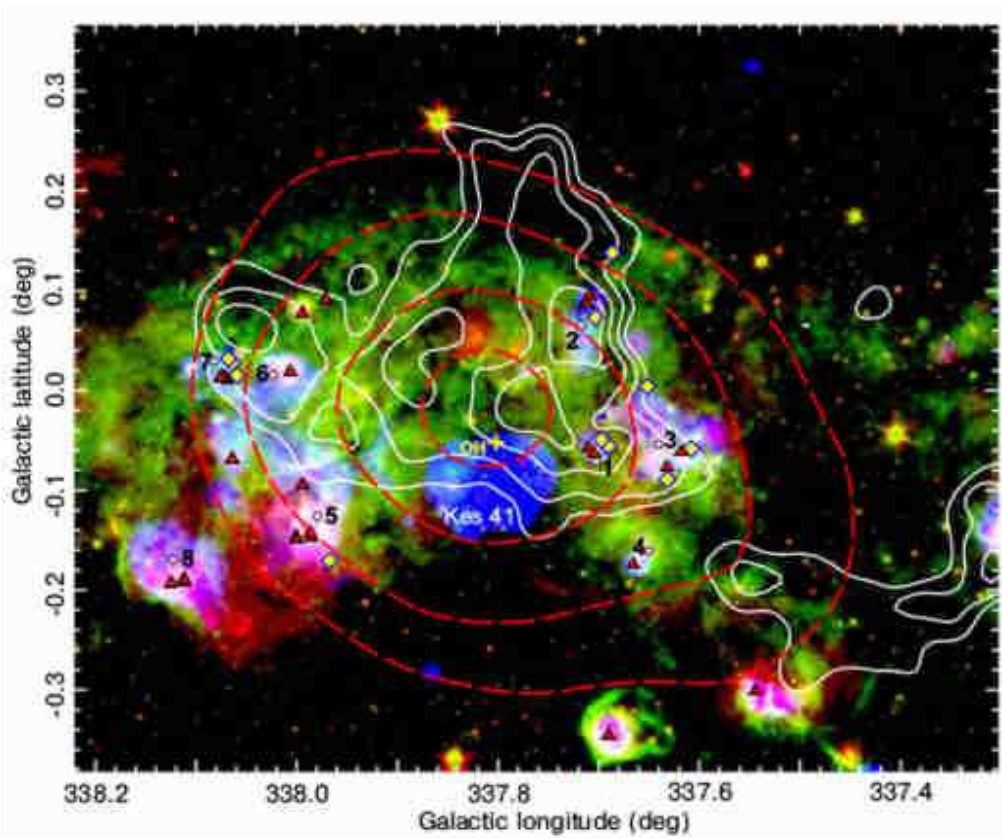}
 \caption{Large-scale three-colour intensity map of the $\gamma$-ray emitting region around SNR~{\kes}: Mid-infrared data at 8 and 24~$\mu$m from MIPSGAL and GLIMPSE {\Spitzer} are displayed in green and red, respectively, while the radio continuum emission at 843~MHz from {\kes} and the HII regions is shown in blue. 
Dashed red contours correspond to statistical significances of 20, 22, 25, and 27$\sigma$ determined from the re-analysis of the $\gamma$-rays photons detected by {\Fermi} in the in the 0.5-400~GeV energy range (Supan et al. 2018b). 
White contours trace the molecular gas emission in the $^{12}$CO ({\COtrans}) line integrated from $-$71 to $-$41~{\kmps} at levels of 87, 101, 119, and 137~K~{\kmps}. Objects are labelled with their names or numbers (see also information in Table~\ref{table_kes41_HIIs}): 
{\kes} and the OH maser \citep{koralesky+98} [in blue]; HII regions [numbers from 1 to 8] \citep{jones+12}; Class~II 6.7~GHz CH$_{3}$OH masers [filled yellow diamonds] \citep{caswell+10_MeOH}; and massive young stellar object candidates [upright filled red triangles] \citep{lumsden13}.}
  \label{field}
\end{figure*}

Figure~\ref{field} displays a three-colour composite map centred at ($l$, $b$)$\simeq$ (\dms{337}{\circ}{8}, \dms{0}{\circ}{0}) corresponding to a large field of about 55{\am}~$\times$~45{\am}  in the direction to SNR~{\kes}. This representation depicts the mid-infrared emission at 8~$\mu$m  and 24~$\mu$m  from {\Spitzer} GLIMPSE and MIPSGAL images \citep{churchwell09, carey09}, along with the radio continuum emission at 843~MHz  detected with a $\sim$1{\am} angular resolution in the 
Sydney University Molonglo Sky Survey \citep[SUMSS,][]{bock+99}. 
The radio- and infrared-emitting objects in the surveyed region are labelled in the figure, and their properties are analysed in this section. The interstellar gas emission observed by Mopra in the $^{12}$CO {\COtrans} line, integrated in the velocity range from $\sim$$-$71 to $-$41~km~s$^{-1}$ (the velocities always refer to that of the local standard of rest, LSR), is traced by white contours (we stress this result again in Sect.~\ref{MC-distance}). In addition, the statistical significance of the GeV emission as revealed by {\Fermi} is overlaid on the infrared and radio image.
The $\gamma$-ray data drawn in Fig.~\ref{field} correspond to an updated analysis of the high-energy emission on the basis of $\text{about nine}$ years of observations with the {\Fermi} telescope. The morphological and spectral modelling from radio- to $\gamma$-ray energies will be investigated in Supan et al. (2018b). In the following, we  focus on the properties of the molecular gas in the large molecular complex reported by the CO Mopra Survey.

\subsection{CO molecular emission}
\label{MC-distance}
\begin{figure*}[!ht]
  \centering
  \includegraphics[width=16cm]{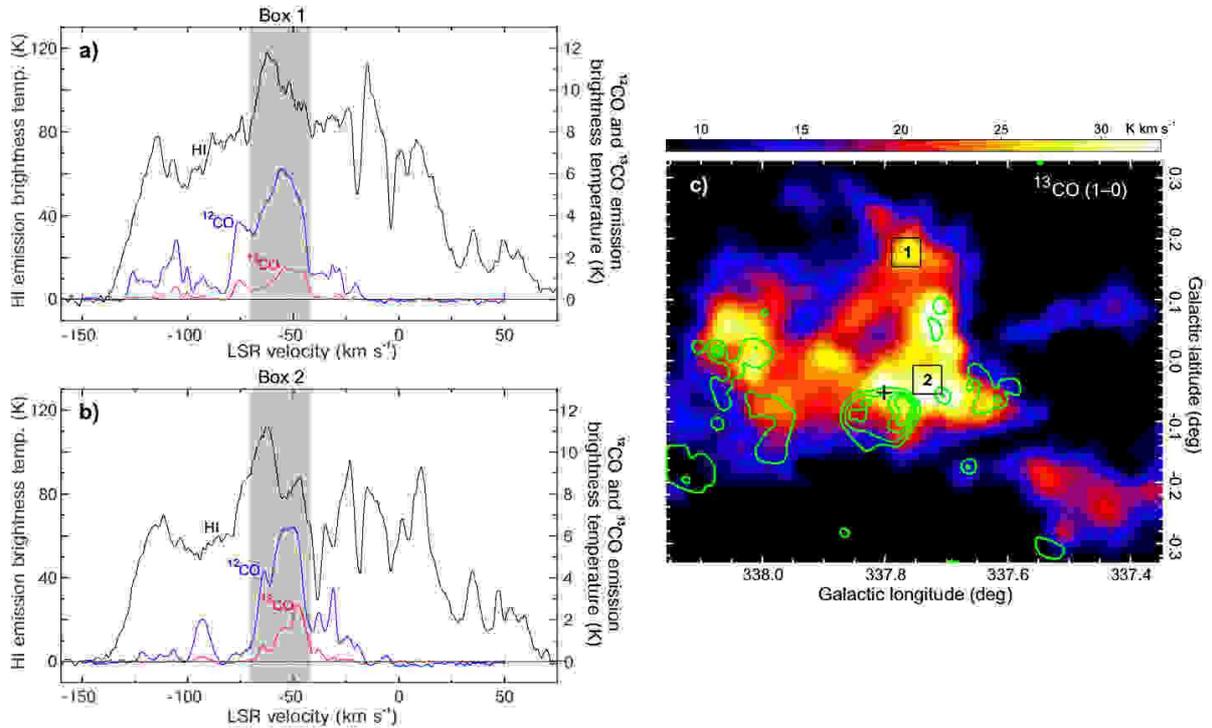}
  \caption{{\bfseries a)}, {\bfseries b)} Test CO and HI 21 cm spectra towards two selected regions, labelled Box 1 and 2, used to determine the distance to the $\gamma$-ray emitting molecular gas. Both areas are depicted in panel {\bfseries c)}. The shaded area corresponds to the LSR velocities in which the molecular emission is dominant. The correspondence between the atomic and molecular emission indicates that the molecular material in Box~1 is located at the far kinematic distance of $\sim$12~kpc. We found similar results (not shown here) across most of the CO distribution corresponding to the large cloud. HI self-absorption inside Box~2 revealed that superposed along the line of sight are a few molecular components that lie in front of the large cloud (see discussion in the text). 
{\bfseries c)} Integrated channel map of $^{13}$CO ({\COtrans}) from the Mopra Survey data over LSR velocities range $-$71 to $-41$~km~s$^{-1}$ (the velocity interval shaded in {\bfseries a)} and {\bfseries b)}). Contours at the 0.09, 0.3, 0.5, and 0.7~Jy~beam$^{-1}$ (in green) from the 843~MHz SUMSS data were included to facilitate the comparison. 
The plus marks the OH maser spot \citep{koralesky+98}.} 
  \label{COHIspectra}
\end{figure*}

\begin{figure*}[!ht]
  \centering
  \includegraphics[width=0.7\textwidth]{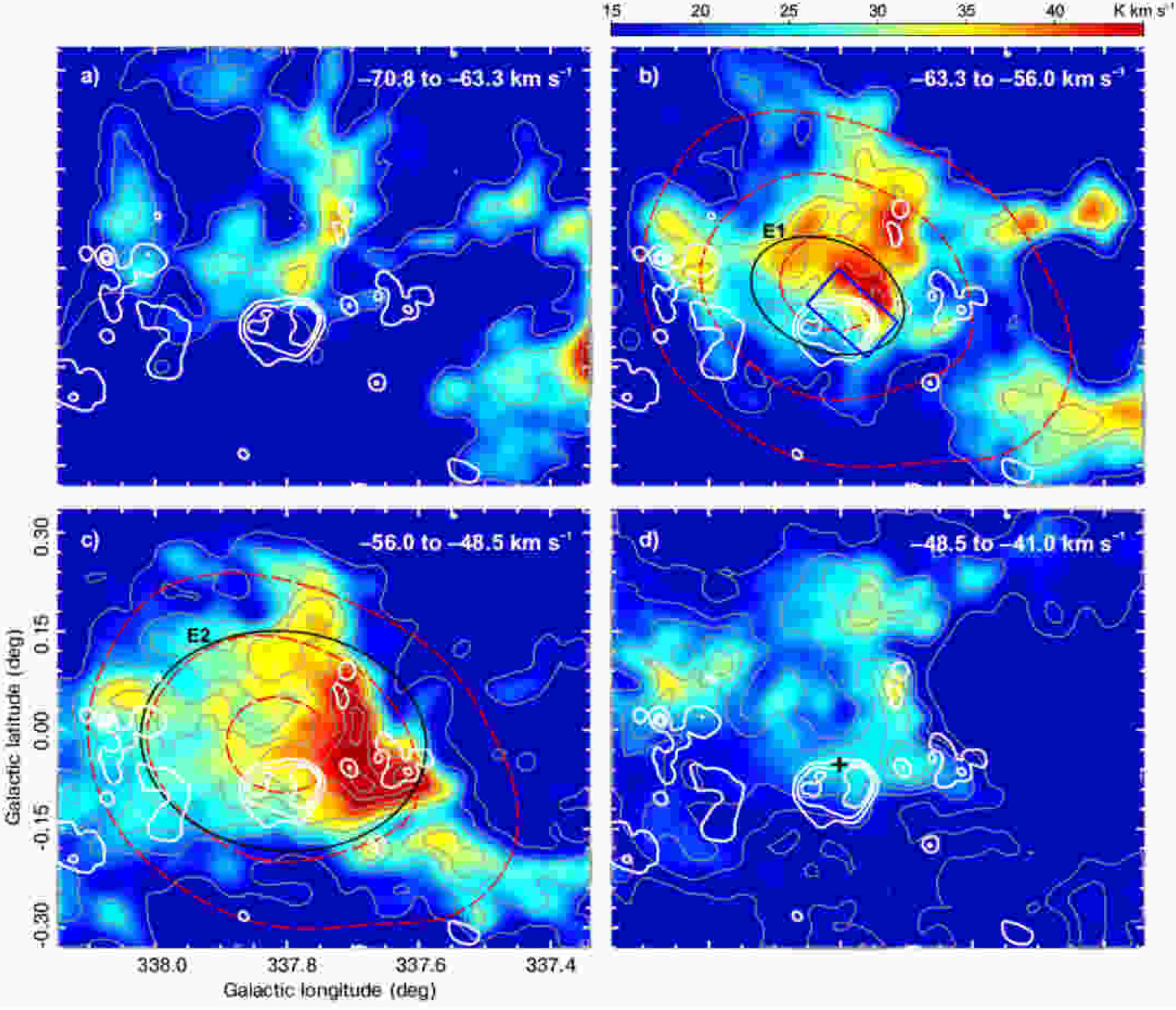}
  \caption{Molecular gas distribution as traced by the $^{12}$CO ({\COtrans}) (in colours) and the $^{13}$CO ({\COtrans}) (in gray contours) data from the Mopra Survey. 
The range of velocities is indicated at the top right corner of the panels. To facilitate the multiwavelength correlations between the remnant {\kes}, the star-forming regions, and the $\gamma$-ray emitting area with their surroundings, the same radio contours (in white) as in Fig.~\ref{COHIspectra} were included in each panel. The $\gamma$-ray emission is drawn as red contours at 20, 24, and 27$\sigma$ (Supan et al. 2018b) only in panels {\bfseries b)} and {\bfseries c)}. 
The areas used in Sect.~\ref{protons} to calculate the ambient properties are represented by the ellipses E1 and E2 included in panels {\bfseries b)} and {\bfseries c)}, respectively. 
The blue rectangle included in panel {\bfseries b)} marks the smaller field studied by \citet{zhang+15}. 
The plus in panel {\bfseries d)} marks the OH maser spot \citep{koralesky+98}.}
  \label{COframes}
\end{figure*}

As a first step, we demonstrate that the molecular gas emission depicted in Fig.~\ref{field} is not the result of a chance coincidence along the line of sight, but corresponds to a single large cloud identified for the first time in the new data collected by the CO Mopra Survey. The hydroxyl (OH) maser emission detected at 1720~MHz is produced between an interacting small cloud and SNR~{\kes} \citep{koralesky+98}, which are both immersed in the large cloud unveiled here by the CO Mopra survey data. 
To establish a reliable distance estimate for all the molecular material, we used data from the CO in the {\COtrans} rotational transition of the $^{12}$CO and $^{13}$CO, in conjunction with spectral-line SGPS observations of the HI. As test cases, Figs.~\ref{COHIspectra}a, b show two sets of CO and HI spectra extracted from two 2$^{\prime}$$\times$2$^{\prime}$ regions, 14$^{\prime}$ and \dms{5}{\prime}{3} away from the maser spot, respectively. These test areas, which we have denoted Box 1 and 2, are overlaid on the integrated intensity image of  $^{13}$CO in Fig.~\ref{COHIspectra}c. 
The depicted $^{13}$CO distribution corresponds to the molecular gas in the $-71$ to $-41$~km~s$^{-1}$ range (the same interval as considered in the display presented in Figs.~\ref{field} and ~\ref{COframes}) in which the molecular emission is prominent. In the spectra corresponding to Box~1, a peak at $\sim$$-$66~km~s$^{-1}$ in the line emission from $^{12}$CO can be discerned within
the surveyed velocity range. 
According to the circular rotation curve model of the Galaxy by \citet{fich+89}, kinematic distances of 4 and 12~kpc are associated with this velocity in the fourth quadrant of the Galaxy. However, the correlation of the CO peak with a maximum in the HI profile permits us to place the interstellar gas in Box~1 at the far kinematic distance of $\sim$12~kpc, which is in broad agreement with the distance calculated to the OH maser emission \citep{koralesky+98}. If the gas within Box~1 were located at the near distance, the cold HI embedded in the molecular gas would absorb the warm HI background emitting at the radial velocity of the CO gas (i.e. $\sim$$-$66~km~s$^{-1}$), and this would be noted by a self-absorption of the HI 21~cm line correlated with the CO emission from the cloud. 
This is in fact what is observed for the gas inside Box~2, for which a valley at $\sim$$-$57~km~s$^{-1}$ in coincidence with a CO peak is clearly visible. 
We interpret this spectral feature as a signature of HI self-absorption, indicating that part of the molecular gas seen in projection inside the CO complex is located at the near position of 4~kpc. It should be also mentioned that the deepest self-absorption line observed for this clump at $-$39~km~s$^{-1}$ was excluded as it lies outside the selected velocity range in which the CO gas was integrated. 
After we inspected the construction of HI and CO spectra over the complete extension of the molecular material shown in Fig.~\ref{COHIspectra}c, we also found weakly pronounced HI valleys that were anti-correlated with the CO line emission between $-$52.5 and $-$48~km~s$^{-1}$. From our analysis, we estimated that in the velocity range in which we are interested ($-$71-$-$41~km~s$^{-1}$), only $\sim$1.5\% of the molecular gas seen in projection as part of the large molecular cloud is placed at the near distance of 4~kpc.

We thus strongly emphasize the importance of the CO Mopra survey in mapping the large-scale structure of the molecular gas. Only in this way are we able to conclude that 
most of the gas with a systematic velocity of $-$71 - $-$41~{\kms} is at a distance of 12.0$\pm$3.6~kpc and forms the natal cloud of the remnant {\kes}, the HII regions, and the massive stellar activity observed in this part of the Galaxy. The main source of error in this estimate is related to uncertainties inherent to the circular rotation model. We quantitatively discuss in Sect.~\ref{protons} the physical conditions of the ambient matter. 

After we determined that a single molecular cloud overlaps (in projection) the $\gamma$-ray emission detected at GeV energies, we proceed to investigate the spatial 
distribution of the CO gas linked to the complex mix of thermal and nonthermal components within it. To do this, we constructed maps corresponding to the {\twelveCO} {\COtrans} rotational line emission integrated every 7.5~km~s$^{-1}$. This is shown in Fig.~\ref{COframes}, where few representative radio continuum and $\gamma$-ray contours are superimposed to facilitate comparison. At the corresponding velocity interval, each
frame also includes grey contours indicating the integrated $^{13}$CO emission. Even though the $^{13}$CO is a better identifier of the denser structures (typically $\sim$10$^{3}$~cm$^{-3}$) than the $^{12}$CO regions, we found that the global distribution of the integrated $^{13}$CO {\COtrans} around the $\gamma$-ray source is similar to that observed for the $^{12}$CO molecular emission in the same transition, where the two gases share peak positions, shapes, and sizes. 

Based on the spatial distribution of the molecular gas, studied in all its extension for the first time in this work, it is evident that the cloud is about 28$^{\prime}$$\times$18$^{\prime}$  (at the estimated distance of 12~kpc, the cloud size is $98 \times 63$~pc). The main part of this cloud lies within the outermost confidence contour of the $\gamma$-ray emission shown in Fig.~\ref{COframes}. 
A local inspection of the cloud shows a bright and roughly asymmetric molecular structure about $\sim$5$^{\prime}$$\times$3$^{\prime}$, which, centred at ($l$, $b$)=(\dms{337}{\circ}{77}, \dms{-0}{\circ}{015}) lies close to the SNR (velocity range between $-$63.3 and $-$48.5~km~s$^{-1}$). 
As outlined above, this region contains the OH maser spot and constitutes a fraction of the molecular material that directly interacts with {\kes}. Additionally, maxima are readily distinguished in the large field of view near ($l$, $b$)=(\dms{337}{\circ}{72}, \dms{-0}{\circ}{065}) and ($l$, $b$)= (\dms{337}{\circ}{73}, \dms{-0}{\circ}{053}) in spatial coincidence with the radio emission from the HII regions in the field (see panels b) and c)). This is an expected result since, as we show below, these correlations correspond to molecular components related to regions of new stellar activity. 
Finally, we did not find any evidence of a shell-like structure in the gas distribution that is morphologically related to SNR~{\kes}, as claimed \citet{zhang+15} based on their analysis of the molecular gas. This discrepancy may arise because the earlier study was performed in a region that only represents a small fraction of the total extent of the cloud shown in Fig.~\ref{COframes}. For comparison purposes, the small rectangle region used by \citet{zhang+15} to calculate the molecular gas parameters is also shown in Fig.~\ref{COframes}b. 

\subsection{Star-forming activity in the GeV-emitting region}
\label{SFR}
In order to 
obtain a comprehensive multiwavelength picture in the region in which we are interested, we also explored the emission at infrared wavelengths since this spectral range is especially adequate to reveal gas structures in their ionised form, such as ionised hydrogen (HII) and star-forming regions.

Figure~\ref{field} clearly shows thermal gas in the field. The correspondence of the emission at mid-infrared wavelengths  with the radio continuum emission is seen as white regions in and around the site where the $\gamma$-ray excess is observed. 
All known HII regions in the Galaxy show this emission distribution, in which the radiation at 24~$\mu$m from the hot dust grains is limited by a polycyclic aromatic hydrocarbon (PAH) region shining at 8~$\mu$m. Eight catalogued HII regions are found to be distributed in the region of the molecular cloud and the $\gamma$-ray source, they are labelled in Fig.~\ref{field} and tabulated in Table~\ref{table_kes41_HIIs} \citep{jones+12}. 
Especially noticeable are the ionised gas regions labelled from 1 to 6 in Fig.~\ref{field}, which shows a clear match with the $\gamma$-ray excess. The HII regions in the field of view, with radio recombination line velocities similar to that of the detected molecular cloud, have measured distances between $\sim$11 and 12.3~kpc, in accord with the distance that we established in Sect.~\ref{MC-distance} for the molecular cloud (see also Table~\ref{table_kes41_HIIs}). 
No infrared counterparts are observed in {\kes}, which is consistent with the nonthermal nature of the radio emission. Nevertheless, it is worth mentioning that although weak, infrared emission at middle wavelengths was detected in several young SNRs, which under restricted conditions originates in the dust grains formed in the SN ejecta \citep{rho08} or by grains heated by the passage of an interacting SN blast wave \citep{williams06}.
\begin{table*}[ht!]
\centering                                                                                                                      
\caption{Parameters of thermal gas and indicators of stellar activity in the field of the {\Fermi} source from available catalogues. The first column indicates the number that labels each HII region as in Fig.~\ref{field}. 
In Columns 2 to 4, the Galactic coordinates (in the $l+b$ form), the radio recombination line (RRL) velocity ($v_\mathrm{RRL}$), and the distance (in kpc) for each HII region are listed. 
The number in Column 5 is the number of 6.7-GHz CH$_3$OH maser detections towards the HII regions, in the velocity range of the MC detected in correlation with the {\Fermi} source. The last column shows the number of MYSO candidates catalogued towards the HII regions. 
``$\geqslant$ TP'' indicates that the HII region is beyond the tangent point ($\sim$8~kpc).}
\label{table_kes41_HIIs}
\begin{tabular}{cccccccc}\hline\hline                                                                                                                   
Source  &  Source position\tablefoottext{a}  &  $v_\mathrm{RRL}$\tablefoottext{a} &  Distance\tablefoottext{a} &   CH$_3$OH                   &  MYSO                        \\
number  &  (Galactic $l+b$)                  &  (K~{\kms})                        &  (kpc)                     &   detection\tablefoottext{b} &  candidates\tablefoottext{c} \hfil\\\hline
        1 & 337.711$-$0.056 &   -50.0 & 12.12           &       2        &      2 \\
        2 & 337.711+0.089   &   -76.7 & 10.79           &       1        &      3 \\
        3 & 337.622$-$0.067 &   -55.0 & 11.84           &       3        &      2 \\
        4 & 337.667$-$0.167 &   -53.0 & 11.95           &      --        &      1 \\
        5 & 337.978$-$0.144 &     --  & $\geqslant$ TP  &       1        &      4 \\
        6 & 338.011+0.022   &   -63.3\tablefoottext{d}  & $\geqslant$ TP & -- & 1 \\
        7 & 337.686+0.137   &   -47.0 & 12.30           &       3        &      3 \\
        8 & 338.114$-$0.193 &   -53.0 & 11.96           &      --        &      4 \\\hline
\end{tabular}
\tablefoot{
\tablefoottext{a}{From \citet{jones+12}.} 
\tablefoottext{b}{From the 6-GHz methanol multibeam maser catalogue of \citet{caswell+10_MeOH}.} 
\tablefoottext{c}{From the Red MSX Source (RMS) Survey \citep{lumsden13}.} 
\tablefoottext{d}{This value corresponds to the LSR velocity of the HII region (not the $v_\mathrm{RRL}$) from {\thirtyCO} observations, according to \citet{anderson+14}.}}
\end{table*}

In the region we are interested in, 6.7~GHz methanol (CH$_{3}$OH) maser members of Class~II were also identified \citep{caswell+10_MeOH}. It is evident that the spatial distribution in the field of the known masers mostly coincides with regions of ionised thermal gas (see also Table~\ref{table_kes41_HIIs}). This identification allows us to infer that high-mass star-forming activity occurs in the region mapped in Fig.~\ref{field} where the $\gamma$-ray flux was detected. It has been demonstrated that this class of methanol maser emission is one of the best markers of massive young stellar objects \citep[see][for further information about the maser classification and its relationship with high-mass young stellar objects]{breen13}. In this respect, we found 16 high-mass protostellar object candidates positionally coincident with the $\gamma$-ray emitting region (see Fig.\ref{field}), although the distances to them are not reported \citep{lumsden13}. 
Notably, in our selected region we did not find catalogued Wolf-Rayet stars,\footnote{See \url{http://www.pacrowther.staff.shef.ac.uk/WRcat} for a compilation of Wolf-Rayet stars.} massive OB associations, or other early-type stars with strong winds \citep{chini12}.

In summary, the infrared-radio correlation together with information from surveys is used here as observational evidence supporting the hypothesis that massive star formation is ongoing in the large molecular cloud.

\subsection{Proton content in the $\gamma$-ray emitting molecular cloud}
\label{protons}
The notorious agreement between the $\gamma$ rays and the CO emission means that it is suspicious that both arise from the same region where interactions of energetic particles (accelerated in a cosmic source) with interstellar matter occur. We determined the properties of the complex environment where {\kes} and the star-forming groups exist by selecting two elliptical zones referred to in the text as ellipses 1 (E1) and 2 (E2), respectively. They are depicted in Fig.~\ref{COframes}. The former, with a major axis of $15^\prime$ and a minor axis of $10^\prime$, comprises the bright molecular region adjacent to SNR~{\kes}. Our selection of a second region reflects the fact that a significant contribution of interstellar material, and hence protons, is also present in a region larger than the portion of the cloud that directly interacts with the remnant. This material spatially corresponds with the star-forming regions observed around the remnant (see Sect.~\ref{SFR}). 
In our analysis of the proton content, we consider the ellipse E2 of axes $26^\prime$ and $20^\prime$ as an appropriate choice encircling all of the radio thermal components. In the subsequent analysis, we determined for each ellipse the proton content based on the contributions from each gas phase, that is, molecular (H$_2$), atomic (HI), and ionised (HII). Therefore, the total proton column density $N_{\mathrm{p}}$ is given by $N_{\mathrm{p}}=N(\mathrm{H_{2}})+N(\mathrm{HI})+N(\mathrm{HII})$. For both areas, we calculated the molecular column density for the $^{12}$CO from Eq.~\ref{W12CO}. For the $^{13}$CO we first derived the optical depth of this gas assuming that the excitation temperatures of 
$^{12}$CO and $^{13}$CO have the same value (Eq.~\ref{tau13}). The obtained optical depth is 0.28. This result indicates that the assumption of an optically thin gas distribution for the $^{13}$CO is valid. Therefore, we used Eq.~\ref{N13CO} to obtain the corresponding molecular column density within the LSR velocity range from $-71$ to $-41$~km~s$^{-1}$. 
Next, we evaluated according to Eq.~\ref{NHI} the atomic contribution to the total column density by integrating the hydrogen gas in both ellipses. 
Additionally, following the procedure described in \citet{katsuta17}, we inferred the number density of ionised hydrogen based on the free-free emission within each elliptical region. We found, however, that the contribution of the ionised gas to the total column density is not significant, corresponding to less than 4\% of the total value for the two regions.

We used the obtained column densities to calculate the proton content $n_{\mathrm{p}}$=$N_{\mathrm{p}}/L$ for each of the considered regions, where we assumed that the thickness $L$ for the interstellar gas along the line of sight equals the average size of each ellipse ($L_{1}$$\sim$43~pc and $L_{2}$$\sim$80~pc at a distance of 12~kpc, for ellipses E1 and E2, respectively). 
On the basis of our results obtained for both isotopologues over each elliptical area, we calculated the total mass of the gas, adopting a distance $d=12$~kpc in the relation, $M=\mu\,m_{\mathrm{H}}\, d^{2}\,\Omega\,N(\mathrm{H_{2}})$, where $\mu$=2.8 is the mean molecular weight if a relative helium abundance of 25\% is assumed, $m_{\mathrm{H}}$ is the hydrogen mass, and $\Omega$ is the solid angle subtended by the region of interest. 
As explained in Sect.~\ref{MC-distance}, we used the self-absorption to resolve foreground CO gas (at a distance of 4~kpc). After a careful inspection of this gas distribution, for ellipse E1 we found a mass of gas of about $1.5 \times 10^4$~{\Msun} that is not related to the large cloud. For ellipse E2, we determined that cold foreground gas lies along the line of sight with a mass approximately equal to $3.4 \times 10^4$~{\Msun}. 
In our measure of the proton content in the SNR surrounding area (E1) and in the larger region encircling all the thermal components in the field (E2), these  masses were not accounted for.

The parameters of the gas in the cloud are listed in Table~\ref{table_kes41_density}. They are typical for giant molecular clouds. The errors in the column density estimates, and consequently, in the determination of masses and number densities, are mainly associated with the selection of the region used to integrate the emission and uncertainties in the determination of the distance. 
We note that our analysis yields physical parameters of the interstellar medium that are consistent with those presented by \citet{zhang+15}, when the differences in the sizes of the inspected regions are taken into account. Thus, we emphasize that the total mass and proton density derived here represent mean values obtained in the large analysed regions inside the cloud, whereas the values ($\sim$20$\times$10$^{4}$~M$_{\odot}$, $n(\mathrm{H_{2}})$$\sim$310-510~cm$^{-3}$) published by \citet{zhang+15} can be considered to be associated with the local gas only around {\kes}. 
In addition, we remark here that the total proton content in the region of {\kes} is similar to that obtained for the interstellar medium towards other well-known interacting medium-age SNRs. Canonical sources that serve as examples of this observation are the remnant RX~J1713.7$-$3946, Vela Jr., HESS~J1731$-$347, and W44 \citep{kuriki+17}, all of which are evolving in environments with high proton densities (hundreds of particles cm$^{-3}$).
\begin{table*}[ht!]
\centering
\caption{Physical properties of the CO, HI, and HII gas.}
\label{table_kes41_density}
\begin{tabular}{cccccccccc}\hline\hline
 \multirow{2}{*}{Region\tablefoottext{a}} & $l$\tablefoottext{b} & $b$\tablefoottext{c} & Size\tablefoottext{d} & \multicolumn{3}{c}{Mean column density [$10^{22}$~\cm{-2}]\tablefoottext{e}}    & Total proton density\tablefoottext{f} & Total mass\tablefoottext{g}\\
           & (deg) & (deg) & (arcmin)       & $N(\mathrm{H_{2}})$    &   $N(\mathrm{HI})$   & $N(\mathrm{HII})$           & $n_{\mathrm{p}}$[\cm{-3}]  & 
$M$[$10^5$~\Msun]   \\\hline
 Ellipse~E1 & $337.82$  & $-0.042$  & 15~$\times$~10 & 5.6~$\pm$~1.3 & 0.44~$\pm$~0.11 & 0.14~$\pm$~0.04  & 950~$\pm$~330 & 9.7~$\pm$~4.0 \\
 Ellipse~E2 & $337.81$  & $-0.017$  & 26~$\times$~20 & 4.9~$\pm$~1.1 & 0.49~$\pm$~0.12 & 0.21~$\pm$~0.05  & 460~$\pm$~160 &  32~$\pm$~14  \hfil\\\hline
\end{tabular}
\tablefoot{
\tablefoottext{a}{Region name. See Fig.~\ref{COframes}.} 
\tablefoottext{b,c}{Galactic longitude and latitude.}
\tablefoottext{d}{Region size.}
\tablefoottext{e}{Mean column densities derived from the CO and the HI 
integrated intensity (in the $-$71-$-$41~km~s$^{-1}$ range), and the ionised gas.}
\tablefoottext{f}{Proton density derived from the mean column densities in Col.~(5) using $N_{\mathrm{p}}=N(\mathrm{H_{2}})+N(\mathrm{HI})+N(\mathrm{HII})$.}
\tablefoottext{g}{Total mass of gas in each selected region.} 
}
\end{table*}

\section{Concluding remarks}
\label{summary}
Using the high-quality data acquired as part of The CO Mopra Southern Galactic Plane CO Survey, we investigated in detail the physical properties of the $^{12}$CO and $^{13}$CO ({\COtrans}) 
gas emission in the direction to SNR~{\kes}. The molecular gas information presented in this work, used in conjunction with neutral atomic hydrogen observations from SGPS and {\Spitzer} mid-infrared data to describe the thermal gas, as well as HII, masers, and young massive stellar objects, provide a complete characterization of the SNR environment. 
Taking advantage of the large area covered by the observations, we uncovered the natal cloud of these objects, which is located at 12.0$\pm$3.6~kpc from us, with a size of $\sim$28$^{\prime} \times$18$^{\prime}$ and covers a broad velocity range from $\sim$$-$71 to $-$41~km~s$^{-1}$. This cloud matches (on the plane of the sky) the $\gamma$-ray emission detected by {\Fermi}. Compared with the previous work of \citet{zhang+15} in a much smaller region only around SNR~{\kes}, this is the first time that the molecular gas towards the $\gamma$-ray emission is analysed in its whole extent. 
For the large cloud, we found a total interstellar proton density of 460$\pm$160~cm$^{-3}$, while for the smaller region enclosing the $\gamma$-ray peak (within the relatively low angular resolution of {\Fermi}) and the molecular material adjacent to {\kes}, the proton density is $950\pm330$~cm$^{-3}$. Both estimates include contributions from the molecular, atomic, and ionised gases in the region.

This work clearly demonstrates the effectiveness of the CO Mopra Survey in the quest of finding large interstellar complexes associated with $\gamma$-ray radiation, which provides high sensitivity and high spatial and spectral resolution. The case discussed here may add to a short list of the few well-known SNRs \citep[e.g. W28, W51C, and IC~443,][]{gab15} that interact with molecular material in massive SFRs. The implications of our analysis on the production of the $\gamma$-ray flux will be investigated in a separate paper (Supan et al. 2018b).

\bibliographystyle{aa}
 \bibliography{mopra}

\begin{acknowledgements}
The authors are grateful to the referee for helpful comments on this manuscript. 
G. Castelletti and A.~D. Supanitsky are members of the {\it Ca\-rre\-ra del Investigador Cient\'{\i}fico} of CONICET, Argentina. L. Supan is a post-doc Fellow of CONICET, Argentina. 
This research was partially supported by grants awarded by ANPCYT (PICT~1759/15) and the University of Buenos Aires (UBACyT 20020150100098BA), Argentina. 
The Mopra radio telescope is part of the Australia Telescope National Facility, which is funded by the Commonwealth of Australia for operation as a National Facility managed by CSIRO. Many staff of the ATNF have contributed to the success of the remote operations at Mopra. 
The University of New South Wales Digital Filter Bank used for the observations with the Mopra Telescope (the UNSW-MOPS) was provided with support from the Australian Research Council (ARC). We also acknowledge ARC support through Discovery Project DP120101585. 
This research has made use of the NASA/IPAC Infrared Science Archive, which is operated by the Jet Propulsion Laboratory, California Institute of Technology, under contract with the National Aeronautics and Space Administration. 
This work is based [in part] on observations made with the {\Spitzer} Space Telescope, which is operated by the Jet Propulsion Laboratory, California Institute of Technology under a contract with NASA.
\end{acknowledgements}

\end{document}